\documentclass[letterpaper]{article} 
\usepackage{aaai23}  
\usepackage{times}  
\usepackage{helvet}  
\usepackage{courier}  
\usepackage[hyphens]{url}  
\usepackage{graphicx} 
\usepackage{natbib}  
\usepackage{caption} 
\usepackage{algorithm}
\usepackage{algorithmic}

\usepackage{newfloat}
\usepackage{listings}
\DeclareCaptionStyle{ruled}{labelfont=normalfont,labelsep=colon,strut=off} 
\floatstyle{ruled}
\newfloat{listing}{tb}{lst}{}
\floatname{listing}{Listing}
\usepackage{booktabs}
\usepackage{multicol}
\usepackage{amsmath}
\usepackage{multirow}
\usepackage{pifont}
\newcommand{\cmark}{\ding{51}}%
\newcommand{\xmark}{\ding{55}}%

\title{Retweet-BERT: Political Leaning Detection Using Language Features and Information Diffusion on Social Networks}
\author{Julie Jiang\textsuperscript{\rm 1, \rm 2}, Xiang Ren\textsuperscript{\rm 1, \rm 2}, Emilio Ferrara\textsuperscript{\rm 1, \rm 2, \rm 3}}
\affiliations{
    \textsuperscript{\rm 1} Information Sciences Institute, University of Southern California, Marina Del Rey, CA, USA \\
    \textsuperscript{\rm 2} Department of Computer Science, Viterbi School of Engineering, University of Southern California, Los Angeles, CA, USA\\
    \textsuperscript{\rm 3} Annenberg School of Communication, University of Southern California, Los Angeles, CA, USA \\ 
    juliej@isi.edu, xiangren@usc.edu, ferrarae@isi.edu}
\begin{document}

\maketitle
\begin{abstract}

Estimating the political leanings of social media users is a challenging and ever more pressing problem given the increase in social media consumption. We introduce Retweet-BERT, a simple and scalable model to estimate the political leanings of Twitter users. Retweet-BERT leverages the retweet network structure and the language used in users' profile descriptions. Our assumptions stem from patterns of networks and linguistics homophily among people who share similar ideologies.  Retweet-BERT demonstrates competitive performance against other state-of-the-art baselines, achieving 96\%-97\% macro-F1 on two recent Twitter datasets (a COVID-19 dataset and a 2020 United States presidential elections dataset). We also perform manual validation to validate the performance of Retweet-BERT on users not in the training data. Finally, in a case study of COVID-19, we illustrate the presence of political echo chambers on Twitter and show that it exists primarily among right-leaning users. Our code is open-sourced and our data is publicly available.

\end{abstract}

\maketitle

\section{Introduction}
Online communities play a central role as the glue of the very fabric of our digital society. This has become even more obvious during the unprecedented times of physical isolation brought by the COVID-19 pandemic, during which social media have seen a significant uptick in engagement \cite{koeze2020virus}. 
Recent work revealed that COVID-19 quickly became a highly politicized and divisive topic of discussion online  \cite{calvillo2020political,jiang2020political}. The latest literature suggests that political affiliations may have an impact on people's favorability of public health preventive measures (e.g., social distancing, wearing masks) \cite{jiang2020political}, vaccine hesitancy \cite{peretti2020future,hornsey2020donald}, and conspiracy theories \cite{uscinski2020people}. Though polarization on social media has been a long-standing phenomenon \cite{conover2011political,colleoni2014echo,an2014sharing,cinelli2020echo}, it is particularly imperative we study how polarization affects the consumption of COVID-19 information.
Divisive politicized discourse can be fueled by the presence of \textit{echo chambers}, where users are mostly exposed to information that well aligns with ideas they already agree with, further reinforcing one's positions due to confirmation bias \cite{garrett2009echo,barbera2015tweeting}. Political polarization can contribute to the emergence of echo chambers \cite{conover2011political,cinelli2020echo}, which may accelerate the spread of misinformation and conspiracies \cite{delvicario2016spreading,shu2017fake,motta2020right,rao2020political,muric2021covid}. 
\begin{figure}
    \centering
    \includegraphics[width=0.9\linewidth]{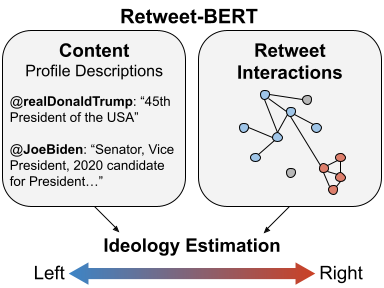}
    \caption{The two key motivating components of Retweet-BERT.}
    \label{fig:rbert_motivation}
\end{figure}
To facilitate research in online polarization, such as  the COVID-19 infodemic, we present \textit{Retweet-BERT}, a lightweight tool to accurately detect user ideology in large Twitter datasets (illustrated in Fig. \ref{fig:rbert_motivation}). Our method simultaneously captures \textit{(i)} semantic features about the user's textual content in their profile descriptions (e.g., affiliations, ideologies, sentiment, and linguistics) and \textit{(ii)} the patterns of diffusion of information -- i.e., the spread of a given message on the social network -- and how they can contribute to the formation of particular network structures (e.g., echo chambers). Prior works on polarization primarily focus on only one of these aspects \cite{conover2011political,conover2011predicting,barbera2015tweeting,preotiuc2017beyond,wong2016quantifying}. 

There are two important assumptions behind Retweet-BERT. One is that the act of retweets implies endorsement \cite{boyd2010tweet}, which further implies support for another's ideology \cite{wong2016quantifying}. The other is that people who share similar ideologies also share similar textual content in their profile descriptions, including not only similar keywords (e.g. \textit{Vote Blue!}) and sentiment, but also linguistics.  The idea of \textit{linguistic homophily} among similar groups of people has been documented and explored in the past \cite{yang2017overcoming,kovacs2020language}. People who adopt similar language styles have a higher likelihood of friendship formation \cite{kovacs2020language}. 

Retweet-BERT leverages both network structure and language cues to predict user ideology. Our method is simple, intuitive, and scalable. The two steps to Retweet-BERT are 
\begin{enumerate}
    \item \textbf{Training} in an unsupervised manner on the full dataset by learning representations based on users' profile descriptions and retweet interactions
    \item \textbf{Fine-tuning} the model for polarity estimation on a smaller labeled subset
\end{enumerate}

\begin{figure}
    \centering
    \includegraphics[width=0.95\linewidth]{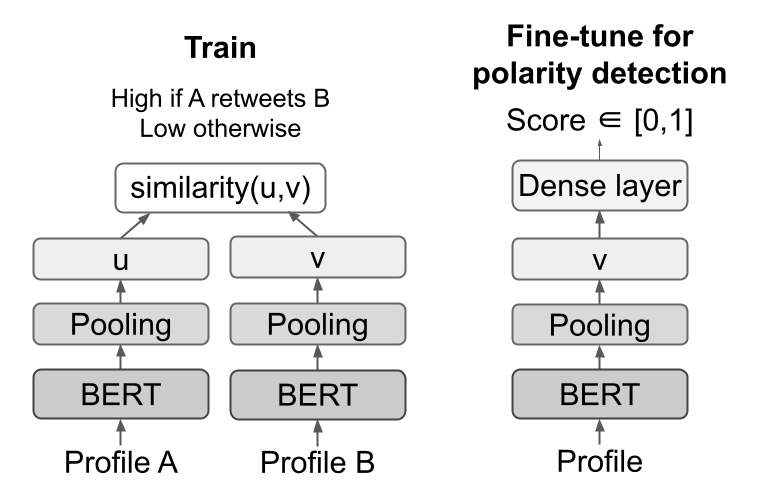}
    \caption{Illustration of the proposed Retweet-BERT. We first train it in an unsupervised manner on the retweet network (left) using a Siamese network structure, where the two BERT networks share weights. We then train a new dense layer on top to predict polarity on a labeled dataset (right).}
    \label{fig:retweetbert}
\end{figure}

An illustration of Retweet-BERT is shown in Fig. \ref{fig:retweetbert}. Crucially, our method does not require human annotations. Instead, we label a small set of users heuristically based on hashtags and mentions of biased new media outlets, as was done in prior works \cite{conover2011predicting,badawy2018analyzing,addawood2019linguistic}. In addition, since we only use profile descriptions instead of all of the users' tweets, Retweet-BERT can be easily deployed.

The datasets we use are two large-scale Twitter datasets collected in recent years. The COVID-19 Twitter dataset was collected from January to July of 2020 for 232,000 active users. We demonstrate that Retweet-BERT attains 96\% cross-validated macro-F1 on this dataset and outperforms other state-of-the-art methods based on transformers, graph embedding, etc. We also perform extensive evaluations of our model on a second Twitter dataset on the 2020 presidential elections to showcase the reliability of Retweet-BERT (97\% macro-F1).

Using Retweet-BERT, we estimate polarity scores for all users in the COVID-19 dataset and characterize patterns of information distribution in a case study COVID-19 on Twitter. Left- and right-leaning users exhibit distinct and asymmetrical patterns of communication. Moreover, we observe a significant presence of echo chambers in the right-leaning population. Our results underscore the urgency and importance of further research in this area.

In sum, the contributions of this work are:
\begin{itemize}
    \item We present Retweet-BERT, a simple and elegant approach to estimate user ideology based on linguistic homophily and social network interactions.
    \item We conduct experiments and manual validations to highlight the effectiveness of Retweet-BERT on two public recent Twitter datasets compared to baselines: COVID-19 and the 2020 US presidential elections.
    \item We illustrate the presence of polarization and political echo chambers on Twitter by applying Retweet-BERT to the COVID-19 dataset.
\end{itemize}
Our code is open-sourced and our data is publicly available through the original dataset papers (see Appendix).


\section{Related Work}
\subsection{Ideology Detection}
There is growing interest in estimating expressed ideologies. Many works focused on opinion mining and stance detection \cite{somasundaran2009recognizing,walker2012stance,abujabara2013identifying,hasan2014taking,sridha2015joint,darwish2020unsupervised}. Of particular interest are political ideology detection of textual data  \cite{sim2013measuring,iyyer2014political,bamman2015open} as well as of Twitter users \cite{conover2011predicting,conover2011political,barbera2015tweeting,yang2016social,wong2016quantifying,preotiuc2017beyond,badawy2018analyzing,badawy2019falls,xiao2020timme}. There are two general strategies for identifying Twitter user ideologies: content-based and network-based. Content-based strategies are concerned with the user's tweets and other textual data. An earlier study used hashtags in tweets to classify users' political ideologies \cite{conover2011predicting}. \citet{preotiuc2017beyond} applied word embedding on tweets to detect tweets of similar topics. Network-based strategies leverage cues from information diffusion to inform ideological differences. These models observe that users interact more with people they share similar ideologies with \cite{yang2016social}. Interactions can be retweets \cite{wong2016quantifying} or followings \cite{barbera2015tweeting}. \citet{xiao2020timme} formulated a multi-relational network using retweets, mentions, likes, and follows to detect binary ideological labels. Other works used a blend of both content- and network-based approaches \cite{badawy2019falls}. Hashtag-based methods were combined with label propagation to infer the leanings of users from the retweet network
\cite{conover2011predicting,conover2011political,badawy2018analyzing}. Closely related to our work, \citet{darwish2020unsupervised} clustered users by projecting them on a space jointly characterized by their tweets, hashtags, and retweeted accounts; however, this algorithm comes at a high computational cost.

\subsection{Socially-infused Text Mining} 
More related to our work is a recent line of work that learns from socially-infused text data. 
\citet{li2019encoding} combined user interactions and user sharing of news media to predict the bias of new articles. \citet{pan2016tri} used node structure, node content, and node labels to learn node representations to classify categories of scientific publications. \citet{yang2017overcoming} used social interactions to improve sentiment detection by leveraging the idea of linguistics homophily. \citet{johnson2017leveraging} used lexical, behavioral, and social information to categorize tweets from politicians into various topics of political issues. These works provide promising results for combining social network data and textual data. 

\paragraph{\normalfont \textbf{Our Work:}} Retweet-BERT is unique from the approaches described above in two substantial ways: (i) it combines both language features, in particular the state-of-the-art transformers (BERT \cite{devlin2019bert}) for natural language processing, and social network features for a more comprehensive estimation of user ideology, and (ii) it is scalable to large datasets without supervision.

\section{Data}\label{sec:data}
We use two recent large-scale Twitter datasets. The primary dataset is on COVID-19 (\texttt{COVID}) from January 21 to July 31, 2020 (v2.7) \cite{chen2020covid}. All tweets collected contain COVID-related keywords. We also use a secondary dataset on the 2020 presidential elections (\texttt{Elections}) collected from March 1 to May 31, 2020 \cite{chen2021election2020}. Both datasets are publicly available. Each tweet contains user metadata, including their profile description, the number of followers, the user-provided location, etc. Users can be verified, which means they are authenticated by Twitter in the interest of the public.

Although a number of Twitter accounts have since been banned by Twitter (notably, @realDonaldTrump was suspended in January 2021 \cite{twitterbantrump}), our data collection was done in real-time and so all tweets by banned accounts are still in our dataset.

\subsection{Content Cues: Profiles} For the purposes of this work, we do not use tweet contents but rather user profile descriptions. In addition to different users posting various numbers of tweets, our main assumption behind this work is that profile descriptions are more descriptive of a user's ideology than tweets. The profile description is a short biography that is displayed prominently when clicking into a user. It usually includes personal descriptors (e.g., ``\textit{Father}'', ``\textit{Governor}'', ``\textit{Best-selling author}'') and, when appropriate, the political ideology or activism they support (e.g., ``\textit{Democratic}'', ``\textit{\#BLM}''). Capped at 160 characters, these descriptions have to be short, which motivates users to convey essential information about themselves clearly, succinctly, and attractively. Previous work established a positive link between the number of followers and the character length of the user \cite{mention2018twitter}, which would suggest that more influential users will have a more meaningful profile.

\subsection{Interaction Cues: Retweet Network} In this work, we use \textit{retweets} to build the interaction network. Retweets refer only to tweets that were shared verbatim. Retweets are distinct from \textit{quoted tweets}, which are essentially retweets with additional comments. We do not use the \textit{following} network as it is rarely used due to the time-consuming nature of its data collection \cite{martha2013study}. The retweet network $G_R$ is a weighted, directed graph where vertices $V$ are users and edges $E$ are retweet connections. An edge $(u,v)\in E$ indicates that user $u$ retweeted from user $v$ and the weight $w(u,v)$ represents the number of retweets.

\subsection{Data Pre-processing}
We removed inactive users and users who are likely not in the U.S. (see Appendix for details). Users in our dataset must have posted more than one tweet.  To remove biases from potential bots infiltrating the dataset \cite{ferrara2020types}, we calculate bot scores using \citet{botometer}, which assigns a score from 0 (likely human) to 1 (likely bots), and remove the top 10\% of users by bot scores as suggested by \citet{ferrara2020types}. 
The \texttt{COVID} dataset contains 232,000 users with 1.4 million retweet interactions. The average degree of the retweet network is 6.15. Around 18k users ($\approx 8\%$) are verified. The \texttt{Elections} dataset contains  are 115,000 users and 3.6 million retweet interactions.

\section{Method} \label{sec:polarity_estimation}
This section describes our proposed method to estimate the polarity of users as a binary classification problem. We first use heuristics-based methods to generate ``pseudo"-labels for two polarized groups of users, which are used as seed users for training and evaluating polarity estimation models. We then introduce several baseline models followed by Retweet-BERT.

\subsection{Pseudo-label Generation} \label{sec:poli_seed_users}
We consider two reliable measures to estimate political leanings for some users, which can be used for model training and automatic, large-scale evaluation. These measures will be used to generate ``pseudo" political leaning labels for a subset of users (i.e., \textit{seed users}). These seed users will be used as the set of training users. 

\subsubsection{Hashtag-based method.} The first method involves annotating the 50 most popular hashtags used in user profiles as left- or right-leaning depending on what political party or candidate they support (or oppose). 17 of these hashtags are classified as left-leaning (e.g. \textit{\#Resist}) and 12 as right-leaning (e.g. \textit{\#MAGA}). The list of hashtags can be found in the Appendix. Users are labeled left-leaning if their profiles contain more left-leaning than right-leaning hashtags and vice versa. We do not consider hashtags appearing in tweets because hashtags in tweets can be used to reply to opposing ideology content  \cite{conover2011political}. Instead, following prior work  \cite{badawy2018analyzing,addawood2019linguistic}, we assume that hashtags appearing in users' self-reported profile descriptions are better indicators of their true ideological affiliations.

\subsubsection{News media-based method.} The second method utilizes media outlets mentioned in users' tweets through mentions or retweets \cite{badawy2019falls,bovet2019influence,ferrara2020characterizing}. Following \citet{ferrara2020characterizing}, we determined 29 prominent media outlets on Twitter. Each media outlet's political bias is evaluated by the non-partisan media watchdog \textsl{AllSides.com} on a scale of 1 to 5 (\textit{left}, \textit{center-left}, \textit{neutral}, \textit{center-right},  \textit{right}). If a user mentions any of these media outlets, either by retweeting the media outlet's Twitter account or by  link sharing, the user is considered to have endorsed that media outlet. Given a user who has given at least two endorsements to any of these media (to avoid those who are not extremely active in news sharing), we calculate their media bias score from the average of the scores of their media outlets. A user is considered left-leaning if their media bias score is equal to or below 2 and right-leaning if their score is equal or above 4.

\subsubsection{Pseudo-labeling seed users.} Using a combination of the profile hashtag method and the media outlet method, we categorized 79,370 ($\approx34\%$ of all) users as either left- or right-leaning. The first, hashtag-based, method alone was only able to label around 16,000 users, while the second, media-based, method labeled around 49,000 users. The two methods overlapped in labeling around 10,000 users. In case of any disagreements between the two methods, which were exceedingly rare at only 200 instances, we defer to the first, hashtag-based method. These users are considered \textit{seed users} for political leaning estimation. 75\% of these seed users are left-leaning, a finding consistent with previous research which revealed that there are more liberal users on Twitter \cite{pew2019sizing}. In our secondary \texttt{Elections} dataset, we tagged 75,301 seed users.

\subsubsection{Pseudo-labeling validation.} This pseudo-labeling method is limited in its capacity for labeling \textit{all} users (\textit{i.e.}, low coverage ratio, covering only 34\% of all users), but it serves as a good starting point for its simplicity. We validated this labeling strategy by annotating 100 randomly sampled users from the main \texttt{COVID} dataset. Two authors independently annotated the data by considering  both the tweets and the profile descriptions to determine the users' political leaning, keeping political neutrality to the extent possible. We then discussed and resolved any annotation differences until reaching a consensus. We attained a substantial inter-annotator agreement (Cohen’s Kappa) of 0.85. 96 users' annotated labels agree  with the pseudo-labels and 4 users' labels cannot be conclusively determined manually. The high agreement with the pseudo-labels makes us highly confident in the precision of our pseudo-label approach. 

\subsection{Methods for Polarity Estimation}\label{sec:lm}
While the pseudo-labels can assign confident political leaning labels for a third of all users, they cannot determine the political leaning of the rest. 
To predict political leanings for all users, we explore several representation learning methods based on users' profile description and/or their retweet interactions. In all of our methods in this section and the one that follows (our proposed method), We do not consider users' tweets. This is because the datasets contain sampled tweets based on keywords and do not encompass any user's full tweets histories. Considering tweets in isolation can bias an algorithm for political leaning detection.

\subsubsection{Word embeddings.}
As baselines, we use pre-trained Word2Vec \cite{mikolov2013distributed} and GloVe \cite{pennington2014glove} word embeddings from Gensim \cite{gensim}. The profile embeddings are formed by averaging the embeddings of the profile tokens.

\subsubsection{Transformers.}
Transformers \cite{devlin2019bert,liu2019roberta,sanh2019distilbert} are state-of-the-art pre-trained language models that have led to significant performance gains across many NLP tasks. We experiment with two different ways to apply transformers for our task: (1) \textit{averaging} the output embeddings of all words in the profile to form profile embeddings, and (2) \textit{fine-tuning} a transformer through the initial token embedding of the sentence (e.g., \texttt{[CLS]} for BERT, \texttt{<s>} for RoBERTa) with a sequence classification head. We use the sequence classification head by \citet{huggingface}, which adds a dense layer on top of the pooled output of the transformer's initial token embedding. 

\begin{table*}[t]
    \centering
    \footnotesize
    
    {
    \begin{tabular}{llcccccccc}
       
        \toprule 
        & & && \multicolumn{3}{c}{\texttt{COVID}} & \multicolumn{3}{c}{\texttt{Elections}}\\
        \cmidrule(lr){5-7} \cmidrule(lr){8-10}
        Model Type &Model  & Profile& Network & Acc. & AUC & F1 & Acc. & AUC & F1\\
        \midrule
        \multirow{2}{*}{\textit{Random and Majority}} &Random & \xmark & \xmark & 0.585 & 0.501 & 0.706 & 0.499 & 0.499 & 0.506\\
        &Majority &\xmark & \xmark &  0.706 & 0.500 & 0.828 & 0.508 & 0.500 & 0.674 \\
        \midrule 
       \multirow{2}{*}{\textit{Average word embeddings}} 
        &Word2Vec-google-news-300 & \cmark& \xmark  & 0.852 & 0.877 & 0.907 & 0.831 & 0.906 & 0.839\\
        & GloVe-wiki-gigaword-300 & \cmark& \xmark  & 0.856 &    0.875 &\textbf{0.909} & 0.835 & 0.908 & \textbf{0.844} \\
        
    \midrule 
    \multirow{5}{*}{\textit{Average transformer output}} 
        &BERT-base-uncased  & \cmark& \xmark & 0.859 & 0.882 & 0.910 &  0.837 & 0.912 & 0.844\\
        &BERT-large-uncased &  \cmark&\xmark & 0.862 & 0.887 & 0.911 & 0.842 & 0.913 & 0.848 \\
        &DistilBERT-uncased &\cmark& \xmark  & 0.863 & 0.888 & 0.912 & 0.845 & 0.919 & 0.851\\
        &RoBERTa-base & \cmark& \xmark  &0.870& 0.898 & 0.917 & 0.853 & 0.925 & \textbf{0.859} \\
        &RoBERTa-large & \cmark& \xmark  &0.882 & 0.914 & \textbf{0.924} & - & - & -\\
    \midrule 
    \multirow{3}{*}{\textit{Fine-tuned transformers}} 
        &BERT-base-uncased  & \cmark&\xmark  &0.900 & 0.932 & \textbf{0.934} &0.902 & 0.963 &\textbf{0.906}\\ 
        &DistilBERT-uncased  & \cmark& \xmark  &0.899 & 0.931 & \textbf{0.934} & 0.899 & 0.962 & 0.904 \\
        &RoBERTa-base & \cmark& \xmark &0.893&0.916 & 0.930 & 0.888 & 0.953 & 0.895\\
    \midrule 
    \multirow{3}{*}{\textit{S-BERT}} 
        &S-BERT-large-uncased & \cmark& \xmark &  0.869 &    0.890&0.916 & 0.849 &0.924 & 0.855 \\
        &S-DistilBERT-uncased  & \cmark& \xmark &0.864 &    0.885 & 0.913 & 0.843 & 0.917 & 0.849\\
        &S-RoBERTa-large &  \cmark& \xmark & 0.879 & 0.903 &\textbf{0.922}  &0.874 & 0.944 & \textbf{0.878}\\

    \midrule 
    \multirow{2}{*}{\textit{Graph embedding}}
        & node2vec* &\xmark& \cmark  & 0.928 & 0.955  &\textbf{0.949} & 0.882 & 0.944 & \textbf{0.883}\\
        & GraphSAGE + RoBERTa-base & \cmark& \cmark  & 0.789 & 0.725 & 0.873 & - & - & -\\
    \midrule
    \multirow{3}{*}{\textit{Retweet-BERT} (our model)} 
        &Retweet-DistilBERT-one-neg &\cmark& \cmark & 0.900  &  0.933 & 0.935 & - & -  & -\\
        &Retweet-DistilBERT-mult-neg  &\cmark&\cmark  &0.935 & 0.965 & \underline{\textbf{0.957}} &  0.973 & 0.984 & \underline{\textbf{0.973}}\\ 
        &Retweet-BERT-base-mult-neg &\cmark & \cmark  & 0.934 & 0.966  & \underline{\textbf{0.957}} & 0.971 & 0.984 & 0.971\\
    \bottomrule
    \multicolumn{10}{l}{*node2vec, a transductive-only model, can only be applied to non-isolated users in the retweet network.}
    \end{tabular}
    }
    \caption{5-fold CV results for political leaning classification on seed users for various models that are tuned via grid-search on the main \texttt{COVID} dataset ($N=79,000$) and the secondary \texttt{Elections} dataset ($N=75,000$). The best F1 (macro) scores for each model type are shown in bold and the best overall scores are underlined. Retweet-BERT outperforms all other models on both datasets.
    }
    
    \label{tab:class_results}
\end{table*}

\subsubsection{S-BERT.}
\citet{reimers2019sbert} proposed Sentence Transformers (S-BERT), which is a Siamese network optimized for sentence-level embeddings. S-BERT outperforms naive transformer-based methods for sentence-based tasks, while massively reducing the time complexity. We directly retrieve profile embeddings for each user using S-BERT's pre-trained model for semantic textual similarity.

\subsubsection{Network-based models.} We explore network-based models such as node2vec \cite{grover2016node2vec}, which learns node embeddings based on structural similarity and homophily, and label propagation, which deterministically propagates labels using the network. Neither of these models can classify isolated nodes in the network. We also experiment with GraphSAGE \cite{hamilton2017inductive}, an inductive graph neural network method that utilizes node attributes to enable predictions for isolated nodes. We use the aforementioned profile embeddings as node attributes.
All profile or network embeddings are subsequently fit with a logistic regression model for the classification task. Hyperparameter-tuning details can be found in the Appendix. The profiles are pre-processed and tokenized according to the instructions for each language model.

With the exception of GraphSAGE, all of these aforementioned methods use either the textual features of the profile description or the network content, but not both. Purely network-based models will do poorly for nodes with only a few connections and may only be suitable for non-isolated nodes. Purely text-based models will do poorly when there are insufficient textual features to inform the models. 

\subsection{Proposed Method: Retweet-BERT}\label{sec:rbert}

\subsubsection{Combining textual and social content.} 
To overcome the aforementioned issues,
we propose Retweet-BERT (Fig.~\ref{fig:retweetbert}), a sentence embedding model that incorporates the retweet network. We base our model on the assumption that users who retweet each other are more likely to share similar ideologies. As such, the intuition of our model is to encourage the profile embeddings to be more similar for users who retweet each other. Retweet-BERT is trained in two steps. The first step involves training in an unsupervised manner on the retweet network, and the second step involves supervised fine-tuning on the labeled dataset for classification. Similar to the training of S-BERT \cite{reimers2019sbert}, the unsupervised training step of Retweet-BERT uses a Siamese network structure. Specifically, using any of the aforementioned models that can produce sentence-level embeddings, we apply it to a profile description to obtain the profile embedding $s_i$ for user $i$. For every positive retweet interaction from user $i$ to $j$ (i.e., $(i,j)\in E$), we optimize the objective
\begin{equation}
    \sum_{k\in V, (i, k)\not\in E}\max(||s_i - s_j|| - ||s_i-s_k|| + \epsilon, 0),
\end{equation}
where $||\cdot||$ is a distance metric and $\epsilon$ is a margin hyperparameter. We follow the default configuration as in S-BERT~\cite{reimers2019sbert}, which uses the Euclidean distance and $\epsilon=1$. We then freeze the learned weights and add a new layer on top to fine-tune on a labeled dataset for classification. 

\subsubsection{Negative sampling.} To optimize the training procedure during the unsupervised training step, we employ negative sampling.  We explore two types of negative sampling strategies. The first is a simple negative sampling (\texttt{one-neg}), in which we randomly sample one other node $k$ for every anchor node in each iteration \cite{mikolov2013distributed}. For simplicity, we assume all nodes are uniformly distributed. The second is multiple negative sampling (\texttt{mult-neg}), in which the negative examples are drawn from all other examples in the same batch \cite{henderson2017efficient}. For instance, if the batch of positive examples are $[(s_{i1}, s_{j1}), (s_{i2}, s_{j2}), ..., (s_{in}, s_{jn})]$, then the negative examples for $(s_{ik}, s_{jk})$, the pair at index $k$, are $\{s_{jk'}\}$ for $k'\in[1,n]$ and $k'\neq k$.

It is worth noting that Retweet-BERT disregards the directionality of the network and only considers the immediate neighbors of all nodes. In practice, we find that doing so balances the trade-off between training complexity and testing performance. Building on the convenience of S-BERT for sentence embeddings, we use the aforementioned S-BERT models pre-trained for semantic textual similarity as the base model for fine-tuning.

\section{Results}\label{sec:characterize_partisan}

We conduct two sets of evaluation to compare the methods: 1) cross-validation over the pseudo-labeled seed users, as an automatic, large-scale evaluation; 2) in-house human evaluation on a set of held-out users, as a complementary evaluation to the first one.  We use the macro-averaged F1 score as the primary metric due to data imbalance. We note that due to our setup, many of the aforementioned related work are not directly comparable. We do not use the following network \cite{barbera2015tweeting,xiao2020timme}. We also do not use manual labeling \cite{wong2016quantifying} or additional external sources to determine user ideology \cite{wong2016quantifying,preotiuc2017beyond}. We do include a comparison with the label propagation method used in \citet{conover2011predicting,conover2011political,badawy2018analyzing} on the held-out users.

Finally, the best model (ours) is selected to classify all the remaining users (non-seed users) to obtain their polarity leaning labels in the \texttt{COVID} dataset. These labels are used to conduct a case study of polarization COVID-19 on Twitter.

\subsection{Automatic Evaluation on Seed Users}
\subsubsection{Baselines.} 
We conduct a 5-fold cross-validation on the seed users (i.e., full set of training users) comparing Retweet-BERT with baselines. In addition, we also use a random  label predictor (based on the distribution of the labels) and a majority label predictor model as additional baselines. 
Table \ref{tab:class_results} shows the cross-validated results for political leaning classification on the seed users, 
Overall, the models perform comparatively similarly between the two datasets. Of all models that do not consider the retweet network, fine-tuned transformers are demonstrably better. Averaging transformer outputs and fine-tuning S-BERTs lead to similar results. For transformers that have a \textit{base} and \textit{large} variant, where the \textit{large} version has roughly twice the number of tunable parameters as the \textit{base}, we see very little added improvement with the \textit{large} version, which may be attributed to having to vastly reduce the batch size due to memory issues, which could hurt performance.\footnote{\url{https://github.com/google-research/bert#out-of-memory-issues}} DistilBERT, a smaller and faster version of BERT, produces comparable or even better results than BERT or RoBERTa. Though the network-based model, node2vec, achieves good performance, it can only be applied on nodes that are not disconnected in the retweet network. While GraphSAGE can be applied to all nodes, it vastly underperforms compared to other models due to its training complexity and time efficiency \cite{wu2020comprehensive}.

Our proposed model, Retweet-BERT, delivers the best results using the DistilBERT base model and the multiple negatives training strategy on both datasets. Other Retweet-BERT variants also achieve good results, which shows our methodology can work robustly with any base language model.

\begin{table}[tb]
    \centering

    \footnotesize
    \begin{tabular}{lccl}
        \toprule 
        Model  & Profile& Network & F1\\ 
        \midrule
        
        RoBERTa-large (\textit{average}) & \cmark& \xmark    & 0.892\\ 
        BERT-base-uncased (\textit{fine-tuned}) & \cmark& \xmark  & 0.908 \\ 
        S-RoBERTA-large (\textit{S-BERT}) & \cmark& \xmark   & 0.909 \\ 
        Label Propagation* & \xmark& \cmark   & 0.910\\ 
        node2vec* & \xmark& \cmark& 0.922 \\ 
        Retweet-BERT-base-mult-neg & \cmark & \cmark   & 0.932 \\ 
        \bottomrule
        
    \end{tabular}
    {*Label propagation and node2vec only predicts labels for nodes connected to the training network (\textit{transductive}), but 10 nodes were not connected and thus were excluded from this evaluation.}
    \caption{Results on 85 users with human-annotated political-leaning labels from a random sample of 100 users without seed labels. Retweet-BERT outperforms all models.
    }
    \label{tab:results_annotated_unlabeled}
   
\end{table}
\subsection{Human Evaluation on Held-out Users} 
For further validation, the authors manually annotated the political leanings of 100 randomly sampled users \textit{without} seed labels. We annotated these users as either left- or right-leaning based on their tweets and their profile descriptions. We were unable to determine the political leanings of 15 people.  We take the best model from each category in Table \ref{tab:class_results} and evaluate them on this labeled set. In this experiment, we also include label-propagation, a simple but efficient method to propagate pseudo-labels through the network commonly used in past work \cite{conover2011predicting,conover2011political,badawy2018analyzing}. The results are reported in Table \ref{tab:results_annotated_unlabeled} for the 85 labeled users. With a macro-F1 of 0.932, Retweet-BERT outperforms all baselines, further strengthening our confidence in our model. 

\begin{figure*}
    \centering
    \includegraphics[width=\linewidth]{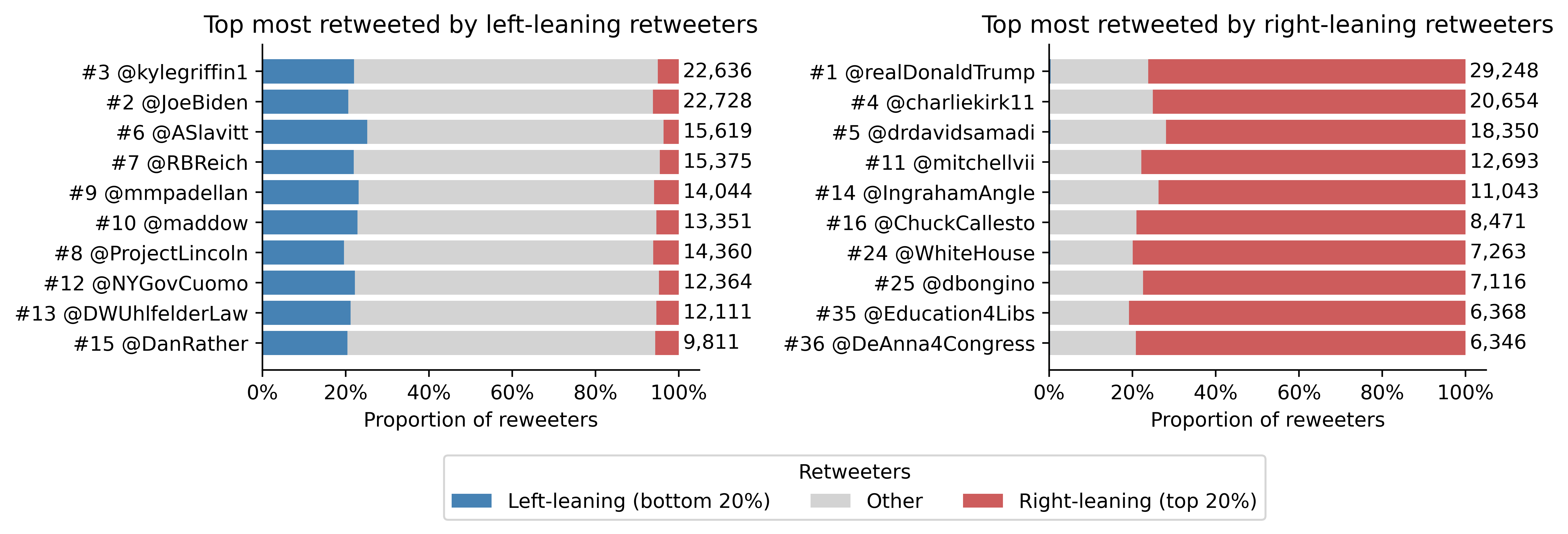}
    \caption{The most retweeted users by the left- and right-leaning user base (reprinted from \citet{jiang2021social}). The bar plots show the distribution of their unique retweeters by political leaning. Users are also ranked by their total number of retweeters (i.e. \#1 @realDonaldTrump means that @realDonaldTrump has the most retweeters overall). Numbers appended to the end of the bars show their total number of retweeters. Accounts most retweeted by left-leaning retweeters are made of 20\% left-leaning retweeters and 5\% right-leaning retweeters, whereas accounts most retweeted by right-leaning retweeters are made of 80\% right-leaning retweeters and virtually no left-leaning retweeters. }
    \label{fig:high_in_degree}
\end{figure*}
\subsection{Case Study of COVID-19}
\label{sec:popular_lr}
To demonstrate the applicability of Retweet-BERT, we apply it to our entire \texttt{COVID} dataset to obtain polarity scores for each user to characterize the extent of polarization online. We reproduce results from our follow-up work, which analyzes the characteristics of partisan users in the \texttt{COVID} dataset.  Here, we use the output sigmoid logits of the Retweet-BERT model, which we interpret as the probability of users being labeled as right-leaning (vs. left-leaning). Since the dataset is imbalanced, we consider only the most likely left-leaning and the most likely right-leaning users of all, which is defined as the top 20\% and bottom 20\% of all users in terms of their predicted polarity scores. We visualize the most retweeted users by the right- and left-leaning user-based and their subsequent audience (retweeters) distribution in Fig. \ref{fig:high_in_degree}, which we break down in detail below. 

\subsubsection{The most popular users among the left the right.} The users depicted in Fig. \ref{fig:high_in_degree} are the users most retweeted by the left- or the right-leaning retweeters.  We use retweet amount as it is a reliable indication of active endorsement \cite{boyd2010tweet} and is also commonly used as a proxy for gauging popularity and virality on Twitter \cite{cha2010measuring}. 


The identities of the top-most retweeted users by partisanship highlight the extent of political polarization. Partisan users mainly retweet from users of their political party. Almost all users who are most retweeted by left-leaning users are Democratic politicians, liberal-leaning pundits, or journalists working for left-leaning media. Notably, @ProjectLincoln is a political action committee formed by Republicans to prevent the re-election of the Republican incumbent President Trump. Similarly, almost all users who are most retweeted by right-leaning users are Republican politicians or right-leaning pundits, or journalists working for right-leaning media. @Education4Libs is a far-right far-right conspiracy group promoting QAnon.\footnote{@Education4Libs is banned by Twitter as of January 2021.}

\subsubsection{Overall popularity of popular users among the left and the right.} These popular users are not only popular among the partisan users, but are considerably popular overall, as indicated by the high overall rankings by the number of total retweeters. With a few exceptions (notably @realDonaldTrump), users who are popular among the left are more popular than users who are popular among the right.

\subsubsection{Audience of the popular users.} Furthermore, we observe a striking discrepancy in the distribution of the audience. The most popular users among the far-right rarely reach an audience that is not also right, whereas those of the far-left reach a much wider audience in terms of polarity, hailing the majority of their audience from non-partisan users (around 75\%)  and, importantly, draw a sizable proportion of far-right audience (around 5\%). In contrast, users who are popular among the far-right have an audience made up almost exclusively of the far-right (around 80\%) and amass only a negligible amount of far-left audience. 

\paragraph*{\normalfont \textbf{Summary:}} Our results highlight that the popular users (i.e., most retweeted) by the left-leaning users are also left-leaning, and vice versa for right-leaning. Additionally, we see that the audience (retweeters) of popular right-leaning users are tweeted almost exclusively by right-leaning users. These results suggest the presence of political echo chambers and asymmetrical information flows between and within the two echo chambers. Additional evaluations of this case study can be found in our follow-up work \citet{jiang2021social}.
\\ 

\section{Discussion}



\subsection{Implications} 

The effectiveness of Retweet-BERT is mainly attributed to the use of both social and textual data. Using both modalities led to significant improvement gains over using only one. This finding has also been validated in other contexts \cite{li2019encoding,pan2016tri,yang2017overcoming,johnson2017leveraging}, but ours is the first to apply this line of thought to detecting user ideology on social media.

Our work can be utilized by researchers to understand the political and ideological landscapes of social media users. For instance, we used it to understand the polarization and the partisan divide of COVID-19 discourse on Twitter. Our results suggest the existence of echo chambers, which warrants further investigation into how political echo chambers may contribute to the spread of politically biased, distorted, or non-factual information. 

Though we apply Retweet-BERT specifically to the retweet network on Twitter, we note that it can be extended to \textit{any} data with a social network structure and textual content, which is essentially any social media. Though we use hashtags as the method to initiate weak labels in place of manual supervision, other methods can be used depending on the social network platform such as user-declared interests in community groups (e.g., Facebook groups, Reddit Subreddits, Youtube channels). We leave investigations of using Retweet-BERT on other social network structures such as following networks and commenting networks for future work. 

\subsection{Limitations}
Since our method relied on mining both user profile descriptions and the retweet network, it was necessary to remove users that did not have profile descriptions or have sufficient retweet interactions (see Appendix). As such, our dataset only contains some of the most active and vocal users. The practical use of our model, consequently, should only be limited to active and vocal users of Twitter. 

Additionally, we acknowledge that Retweet-BERT is most accurate on datasets of polarizing topics where users can be distinguished almost explicitly through verbal cues. This is driven by two reasons. First, polarizing datasets makes it clearer to evaluate detection performance. Second, and more importantly, the applications of Retweet-BERT are realistically more useful when applied to controversial or polarizing topics. Since our detection method relies on users revealing explicit cues for their political preference in their profile descriptions or their retweet activities, we focus on the top 20\% (most likely right-leaning) and the bottom 20\% (most likely left-leaning) when conducting the case study on the polarization of COVID-19 discussions. The decision to leave most users out is \textit{intentional}: we only want to compare users for which Retweet-BERT is most confident in predicting political bias. Detecting user ideology is a difficult and largely ambiguous problem, even for humans \cite{elfardy2016addressing}. \citet{cohen2013classifying} raised concerns that it is harder to predict the political leanings of the general Twitter public, who are much more ``modest'' in vocalizing their political opinions. Thus, we focus our efforts on detecting the more extreme cases of political bias in an effort to reduce false positives (predicting users as politically biased when in fact they are neutral) over false negatives (predicting users as politically neutral when in fact they are biased). 



\section{Conclusion}
We propose Retweet-BERT, a simple and elegant method to estimate user political leanings based on social network interactions (the social) and linguistic homophily (the textual). We evaluate our model on two recent Twitter datasets and compare it with other state-of-the-art baselines to show that Retweet-BERT achieves highly competitive performance (96\%-97\% macro-F1 scores). Our experiments demonstrate the importance of including both the textual and the social components. Additionally, we propose a modeling pipeline that does not require manual annotation, but only a training set of users labeled heuristically through hashtags and news media mentions.  Applying Retweet-BERT to users involved in COVID-19 discussions on Twitter in the US, we find strong evidence of echo chambers and political polarization, particularly among the right-leaning population. Importantly, our work has the potential to advance future research in studying political leanings and ideology differences on social media.

\appendix
\section{Appendix}
\subsection{Reproducibility}
\subsubsection{Code and Data Availability.}
We uploaded the code of Retweet-BERT to \url{https://github.com/julie-jiang/retweet-bert}. In accordance with Twitter data-sharing policies, we cannot release the actual tweets. To reproduce our work, the tweets need to be hydrated (see \url{https://github.com/echen102/COVID-19-TweetIDs}) to obtain the profile descriptions of users and to build the retweet network. 

\subsubsection{Heuristics-based Pseudo-Labeling Details.}
We show the exact hashtags in Table \ref{tab:top_50_hashtags} and media bias ratings in Table \ref{tab:media_all_sides} used in the heuristics-based pseudo-labeling of user political leanings. In the labeling process, all hashtags are treated as case insensitive.

\begin{table}
    \centering

    \begin{tabular}{ll}
        \toprule
         \textbf{Left}& \textbf{Right} \\
         \midrule
         Resist & MAGA \\
         FBR & KAG \\
         TheResistance & Trump2020\\
         Resistance & WWG1WGA\\
         Biden2020 & QAnon \\
         VoteBlue & Trump \\
         VoteBlueNoMatterWho & KAG2020\\
         Bernie2020 & Conservative \\
         BlueWave & BuildTheWall \\
         BackTheBlue & AmericaFirst \\
         NotMyPresident & TheGreatAwakening \\
         NeverTrump & TrumpTrain \\
         Resister &\\
         VoteBlue2020\\
         ImpeachTrump \\
         BlueWave2020\\
         YangGang\\
         \bottomrule         
    \end{tabular}
    \caption{Hashtags that are categorized as either left-leaning or right-leaning from the top 50 most popular hashtags used in user profile descriptions in the \texttt{COVID} dataset. }
    \label{tab:top_50_hashtags}
\end{table}
\begin{table}[h]
    \begin{tabular}{llll}
        \toprule
        \textbf{Media (Twitter)} & \textbf{URL} & \textbf{Rating} \\
        \midrule
        @ABC & abcnews.go.com & 2 \\
        @BBCWorld & bbc.com & 3\\
        @BreitbartNews & breitbart.com & 5 \\
        @BostonGlobe & bostonglobe.com & 2 \\
        @businessinsider & businessinsider.com & 3\\
        @BuzzFeedNews & buzzfeednews.com & 1\\
        @CBSNews & cbsnews.com & 2 \\
        @chicagotribune & chicagotribune.com & 3\\
        @CNBC & cnbc.com & 3 \\
        @ CNN & cnn.com & 2 \\
        @DailyCaller & dailycaller.com & 5\\
        @DailyMail & dailymail.co.uk & 5 \\
        @FoxNews & foxnews.com & 4 \\
        @HuffPost & huffpost.com & 1 \\
        InfoWars* & infowars.com & 5\\
        @latimes & latimes.com & 2 \\
        @MSNBC & msnbc.om & 1 \\
        @NBCNews & nbcnews.com & 2 \\
        @nytimes & nytimes.com & 2 \\
        @NPR & npr.org & 3 \\
        @OANN & oann.com & 4\\
        @PBS & pbs.org & 3\\
        @Reuters & reuters.com & 3 \\
        @guardian & theguardian.com & 2 \\
        @USATODAY & usatoday.com & 3 \\
        @YahooNews & yahoo.com & 2 \\
        @VICE & vice.com & 1 \\
        @washingtonpost & washingtonpost.com & 2 \\
        @WSJ & wsj.com & 3\\
        \bottomrule
    \end{tabular}
    {*The official Twitter account of InfoWars was banned in 2018.}
    \caption{The Twitter handles, media URL, and bias ratings from AllSides.com for the popular media on Twitter.}
    \label{tab:media_all_sides}
\end{table}
\subsubsection{Data Pre-processing.}
We restrict our attention to users who are likely in the United States, as determined by their self-provided location \cite{jiang2020political}. Following \citet{garimella2018quantifying}, we only retain edges in the retweet network with weights of at least 2. Since retweets often imply endorsement \cite{boyd2010tweet}, a user retweeting another user more than once would imply a stronger endorsement and produce more reliable results. As our analyses depend on user profiles, we remove users with no profile data. We also remove users with degrees less than 10 (in- or out-degrees) in the retweet network, as these are mostly inactive Twitter users.

\subsubsection{Hyperparameter Tuning}
All models producing user (profile and/or network) embeddings are fit with a logistic regression model for classification. We search over parameter \{\texttt{C}: [1, 10, 100, 1000]\} to find the best 5-fold CV value. We also use randomized grid search to tune the base models. For node2vec, the search grid is \{\texttt{d}: [128, 256, 512, 768], \texttt{l}: [5, 10, 20, 80], \texttt{r}: [2, 5, 10], \texttt{k}: [10, 5], \texttt{p}: [0.25, 0.5, 1, 2, 4], \texttt{q}: [0.25, 0.5, 1, 2, 4]\}. For GraphSAGE, the search grid is \{activation: [\texttt{relu}, \texttt{sigmoid}], $S_1$: [10, 25, 50], $S_2$: [5, 10, 20], negative samples: [5, 10, 20]\}. Both node2vec and GraphSAGE are trained for 10 epochs with hidden dimensions fixed to 128. Retweet-BERT is trained for 1 epoch. 
\section{Ethical Statement}
We believe our work has the potential to be used in combating misinformation and conspiracy spread, as well as identifying  communication patterns between and within polarized communities. However, we are aware of the ways our work can be misused. For instance, malicious actors can use our work to politically target certain groups of users and propagate the spread of misinformation. As such, we encourage researchers to use these tools in a way that is beneficial for society. Further, to protect the privacy of the users and also in accordance with Twitter's data sharing policy, we will not be sharing our actual dataset, nor the partisan labels, but only the Tweet IDs used in this paper through the original dataset release papers \cite{chen2020covid,chen2021election2020}. All data used in this paper are public and registered as IRB-exempt by the University Southern California IRB (approved protocol UP-17-00610).

\section{Acknowledgments} The authors are grateful to DARPA (award number HR001121C0169) for its support. 
{
\bibliography{mybib}

}
\end{document}